\documentclass[12pt]{article}
\usepackage[T1]{fontenc}
\usepackage[utf8]{inputenc}
\usepackage{fancyhdr}
\usepackage[dvips]{graphicx}
\usepackage{graphicx}
\usepackage{color}
\usepackage{multicol}
\usepackage{float}
\usepackage{makeidx}
\usepackage{setspace}
\usepackage[sort&compress, numbers]{natbib}
\author{Mateusz Denys\footnote{email: mateusz.denys@fuw.edu.pl},~ Tomasz Gubiec, Ryszard Kutner}

\title{Reinterpretation of Sieczka-Hołyst financial market model}

\date{}

\begin{document}

\maketitle

\begin{center}
\emph{Faculty of Physics, University of Warsaw\\
Hoża 69, PL-00681 Warsaw, Poland}
\end{center}

\begin{abstract}
In this work we essentially reinterpreted the Sieczka-Hołyst (SH) model to make it more suited for description of real markets. For instance, this reinterpretation made it possible to consider agents as crafty. These agents encourage their neighbors to buy some stocks if agents have an opportunity to sell these stocks. Also, agents encourage them to sell some stocks if agents have an opposite opportunity. Furthermore, in our interpretation price changes respond only to the agents' opinions change. This kind of respond protects the stock market dynamics against the paradox (present in the SH model), where all agents e.g. buy stocks while the corresponding prices remain unchanged. In this work we found circumstances, where distributions of returns (obtained for quite different time scales) either obey power-law or have at least fat tails. We obtained these distributions from numerical simulations performed in the frame of our approach.
\end{abstract}


\section{Introduction}

The market modeling is a modern challenge, especially in the context of still lasting a global financial crisis. Besides, the market modeling is a canonical branch of econophysics \cite{BFL, TG, JAL}. One of the significant part of this branch is agent based modeling inspired by Ising and Potts models. Notably, the approach based on the Potts model and its different generalizations are particularly fruitful \cite{Bornholdt, BornholdtInni, Iori, LM, Sornette, ZS, Denys}. On this way already several, although not all stylized facts, are well reproduced, e.g. fat tails of returns' distribution as well as long-range correlations of absolute returns, and short-range correlated returns. Certainly, this is a significant encourage to deal with these mo\-dels. Nevertheless, since two decades of fast development of econophysics, there is still no definite model reproducing \emph{all} known up to now stylized facts concerning real financial markets. Furthermore, even a micro\-economic (or microscopic) sources of some stylized facts are still not well understood.

In this paper we mainly reinterpret the Sieczka-Hołyst (SH) threshold model of financial markets \cite{SH}, taking into account the concept of the negotiation round from the Iori model \cite{Iori}. We choose the SH model for the following reasons:
\begin{itemize}
\item[(i)] It contains a threshold mechanism with three possible values of spin variable.
\item[(ii)] It considers interaction between agents as well as noise in their activity.
\item[(iii)] It assumes strength of the interaction which strongly depends on the macroscopic state of the system.
\item[(iv)] The SH model seems to be sensitive to the concrete kind of emotions which each agent subordinates during the stock market evolution \cite{CCH, KD}
\item[(v)] The SH model seems to be a more realistic than other agent-based models.
\end{itemize}
Despite of these properties, the SH model leads to the paradoxal situation where if all agents buy stocks or all sell them, the corresponding prices remain still unchanged. This paradox inspired us for the appropriate reinterpretation of the SH model.

Moreover, in our approach we model emotions of agents, including even very intensive ones, by the mean of the Weierstrass-Mandelbrot noise \cite{RK} instead of the Gaussian noise considered in the SH model.

\section{Sketch of the SH model}
\label{SH}

Let us shortly remind the SH model. There are $N$ interacting agents placed on 2-dimensional square lattice $n \times n$, where $N = n^2$. Agent number $i$ (where $i=1,\ldots,N$) can take into account only one of the three actions: buy, stay inactive or sell. Such an agent is represented by 3-state spin variable $s_i$ with possible values: $+1$ if agent is buying stocks, $0$ if he stays inactive, and $-1$ in the case of selling stocks.

The single time step $t$, called a round, consists of $N$ drawings of spins\footnote{Single round resembles 1MCS/spin used in dynamic Monte Carlo methods.}. Hence, in a given round each agent has, in average, a single chance to change his opinion. After each drawing the chosen spin is updated according to the rule
\begin{equation}
s_i(t)=\mathrm{sign}_{\lambda |M(t-1)|}\left[\sum_{j=1}^{N}J_{ij}s_{j}(t-1)+\sigma \eta_i(t)\right],
\label{s_i(t)}
\end{equation}
where
\begin{equation}
\textrm{sign}_{q}(x)=\left\{ \begin{array}{lll}
+1 & \textrm{if} & x \geq q,\\
0 & \textrm{if} & -q \leq x < q,\\
-1 & \textrm{if} & x < -q,
\end{array} \right.
\label{sign}
\end{equation}
and constants $\lambda, \sigma$ are positive. The value of pair interaction strength $J_{ij}=J$ if agent $j$ is one of the four nearest neighbors of agent $i$, otherwise $J_{ij}=0$. The temporal noise $\eta_i(t)$ is a random variable drawn from the standard normal distribution. It represents individual erratic opinion of $i$-th agent. Furthermore, the change of any spin can affect its neighbors immediately (i.e. within the same round).

The magnetization of the network, present in Expression (\ref{s_i(t)}) is defined as usual,
\begin{equation}
M(t)=\frac{1}{N}\sum_{i=1}^{N} s_{i}(t).
\label{MN}
\end{equation}
The crucial role of the magnetization, as a quantity determining activity of agents, is well seen. That is, index $\lambda |M(t-1)|$ in the sign function is a temporal threshold. Hence, if the absolute value of the neighbors' impact on the $i$-th agent together with his (erratic) opinion is lower than the threshold, the agent does not participate in the market game assuming a neutral position (as his spin variable equals $0$). In other words, the large absolute value of magnetization gives (at a fixed value of $\lambda$) the large value of the threshold. Hence, the probability that each agent assumes a neutral position distinctly increases.

As the value of magnetization was already defined, the temporal price of a stock
\begin{equation}
P(t) \propto \exp{(M(t))}.
\label{P(t)}
\end{equation}

In the simulation Sieczka and Hołyst used a square lattice of agents with $n = 32$ and $N = 1024$, possessing the periodic boundary conditions, and assuming the initial configuration of spins as random. For thermalization of the system they used the first 5000 rounds.

As authors claimed, simulations' results reproduced main stylized facts such as volatility clustering in the variograms of returns, fat (but not the desired power-law) tails of returns' statistics and fast decaying auto\-correlation function of returns revealing a negative feedback. However, autocorrelation of absolute returns decayed exponentially and not according to the desired power-law. In the frame of our approach we repaired both the statistics of returns and autocorrelation of absolute returns.

\section{Reinterpretation of the SH model}
\label{reint}

In this Section, we reinterpret the SH model by essential change of the meaning of the spin variable $s_i \ (i=1,\ldots,N)$. We postulate that instead of an action of the agent, this variable means only the agent's \emph{opinion} about the market situation, which he communicates to his nearest neighbors in accordance with Expression (\ref{s_i(t)}). That is, $s_i = +1$ means a positive opinion communicated by the agent (i.e. according to this message his nearest neighbors should buy stocks). For $s_i=-1$ a negative opinion is communicated by him (i.e. his nearest neighbors should sell stocks). Value $s_i = 0$ simply means a lack of the agent opinion.

In our approach the activity of the agent always requires two subsequent opinion stages. That is, the change of spin variable $d_i(t)=s_i(t)-s_i(t-1)$ (i.e. the change of the agent's opinion during the subsequent rounds) means, for $d_i>0$, the agent's demand and reversely, for $d_i<0$, this change means the agent's supply (considered here as a negative demand).

In other words, the agent declares demand for stocks if his spin variable increased within a given round (in comparison with value of the spin variable obtained within the previous round), while he offers a supply of stocks if the value of his spin variable decreased. For instance, the change of the spin variable from $-1$ to $+1$ (or from $+1$ to $-1$) defines the largest possible demand (or supply) declared by a given agent in a single round. This is because the absolute value of the difference between subsequent spin values is the largest one, i.e. $|d_i(t)| = 2$. Other changes define the smaller demand or supply as then $|d_i(t)| = 1$. We assume that declared demand or supply of any agent is realized immediately, i.e. in the same round. As a result, the agent buys or sells stocks. Note that in this approach the agent opinion concerning the market situation always remains unchanged (by definition) until the next transaction will occur. Thus, two opinion stages (i.e. two subsequent different values of agent's spin) were realized.

Obviously, after buying some stocks the agent immediately (by definition; cf. Expression (\ref{s_i(t)})) communicates the optimistic or neutral opinion to his neighbors, as he is interested in an increase of a stock price. In the opposite case (i.e. in the case of selling), the agent gives a pessimistic or neutral opinion to his neighbors (as he is interested in a decrease of a stock price directly after the selling). This is the way how the subsequent two-stage cycle of the agent activity (i.e. buying or selling and communication) is repeating.

One of the consequences of our reinterpretation of spin variable is that the agents are assumed crafty. That is, if the spin value of agent $s_i=+1$, he communicates positive opinion to his neighbors but he cannot buy a stock by himself. It is because his spin value cannot increase, i.e. he can only keep or sell stocks. For $s_i=-1$ the situation is reversed, i.e. the agent which earlier sold some stocks, now can only keep or buy them. In this way, the negative coupling between subsequent single agent's opinions can be realized. For $s_i=0$ the agent occupies a neutral position and he can either buy or sell a stock as well. Indeed, in this paragraph the principal differences between our approach and the SH model were precisely defined.

In our approach the noise term $\sigma \eta_i(t)$ indicates an intrinsic (erratic) opi\-nion of agent $i$ and parameter $\sigma$ is a strength of this opinion. It should not be mixed up with the spin variable $s_i(t)$, which represents opinion directly communicated to other agents.

Our reinterpretation of the spin variable protects the stock market dynamics against the situation (present in the SH model), where all agents buy stocks while corresponding temporal prices remain unchanged (as magnetization becomes unchanged). Furthermore, our approach ensures correlation between volatility of returns and a trading volume, as expected.

As another significant consequence of our approach, the temporal threshold is able to slow down both the fast increase of the price if trend is positive and its fast decrease if trend is negative. That is, the threshold introduces a kind of a price damping, which cannot lead neither to oscillations nor to change of trend direction.

Usually, in agent-based models one considers simplified linear model of a price formation, where change of price (or logarithmic return) is proportional to excess demand \cite{RJB}. Namely,
\begin{equation}
\ln{P(t)} - \ln{P(t-\tau)} \stackrel{\rm def.}{=} r_{\tau}(t) \propto ED_{\tau}(t).
\label{ror}
\end{equation}
Hence, for the particular case $\tau = 1$ the excess demand can be written in the following form,
\begin{equation}
ED(t) = \sum_{i=1}^{N}d_{i}(t) = \sum_{i=1}^{N}(s_{i}(t) - s_{i}(t-1)).
\label{ED}
\end{equation}
Although Expression (\ref{ror}) is formally equivalent to Equation (\ref{P(t)}), it has a different interpretation as spin $s_i(t)$ means here an opinion of $i$-th agent and not his individual demand (as it is in the SH model). Apparently, the excess demand, $ED(t)$, can change if and only if the opinion of any agent changes. Notably, in the SH model the excess demand does not change although an agent still buys a stock. Due to our reinterpretation of a spin variable, we avoided this paradox.

In the SH model the noise is significant only to some extent. That is, if the value of the noise exceeds the range $[-(4J + \lambda), 4J + \lambda]$, the exact value of this noise is then unimportant. This comes from the threshold character of Expression (\ref{s_i(t)}).

In Expression (\ref{s_i(t)}), there are present three parameters which (without loss of generality) can be reduced to two crucial relative parameters:
\begin{itemize}
\item[(i)] parameter $\alpha_{\lambda} = \frac{\lambda}{4J}$. If $\alpha_{\lambda} < 1$, then the nearest neighbors' impact on the agent is high, otherwise the agent is an opportunist or skeptical as he often chooses $s_i = 0$.
\item[(ii)] Parameter $\alpha_{\sigma} = \frac{\sigma}{4J}$. If $\alpha_{\sigma} < 1$, the agent trusts more his neighbors than himself, otherwise he trusts more himself than his neighbors.
\end{itemize}
In paper \cite{SH} it was considered, for instance, the case $\alpha_{\lambda} > 1$ and $\alpha_{\sigma} < 1$, i.e. opportunist or skeptical agents, trusting more their neighbors than themselves. If agents are opportunist or skeptical, then even coherent feedback coming from all neighbors (having the same value of spin variable) can leave the agent's opinion unchanged. It seems to be useful to systematically examine all ranges of parameters $\alpha_{\lambda}$ and $\alpha_{\sigma}$.

\section{Results and discussion}
\label{results}

We begin our simulation with randomly oriented spins distributed over the square lattice consisting of $N = 1024$ sites. In our simulation we used noise distribution in the form of the Weierstrass-Mandelbrot probability density function,
\begin{equation}
p(x) = \left(1 - \frac{1}{K} \right) \sum_{j=0}^{\infty} \frac{1}{K^j}\cdot\frac{1}{2}\delta(|x| - b_0 b^j), \ K, b > 1, \ b_0 > 0.
\label{WM}
\end{equation}
We used this distribution for the following reasons:
\begin{itemize}
\item[(i)] Agent's emotions, which can contain even very intensive ones, should be ruled by power-law distributions which may be better suited than canonical Gaussian ones.
\item[(ii)] Herein, the control of the noise by its variance and exponent is possible. This variance can vary in the range from finite to infinite values.
\end{itemize}

The Weierstrass-Mandelbrot distribution is a canonic simple one possesing above given properties. This distribution is discrete having spikes at $x = \pm b_0 b^j, \; j = 0, 1, 2, \ldots$. Its variance is finite only for $b^2 / K < 1$ and takes the form
\begin{equation}
\sigma^2 = b_0^2 \frac{1 - 1/K}{1 - b^2/K}.
\label{sigma2}
\end{equation}
Otherwise, it is an infinite quantity.

It can be easily proved that for $|x| \gg 1 / \ln{b}$ Expression (\ref{WM}) can be approximated by
\begin{equation}
p(x) = \frac{1-1/K}{\ln{K}} \frac{b_0^{\beta}}{|x|^{1+\beta}}, \ \textrm{where exponent} \ \beta = \frac{\ln{K}}{\ln{b}}.
\label{appWM}
\end{equation}
Note that $\sigma^2$ is finite only for $\beta > 2$ otherwise, it is infinite.

Although Equation (\ref{appWM}) is a power-law distribution, only the part of this distribution extended over the maximal range $[-(4J+\lambda),4J+\lambda]$ is significant for the system dynamics (as it is suggested by Expression (\ref{s_i(t)})). Inside this maximal range spin can assume values equal to $-1, 0, +1$. Above the right border of this range only value of spin equal to $+1$ appears while below the left border only $-1$ appears. How this affects the the power-law tail of the noise distribution is still an open question.

\begin{figure}[htb]
\begin{center}
\includegraphics[width=0.8\textwidth]{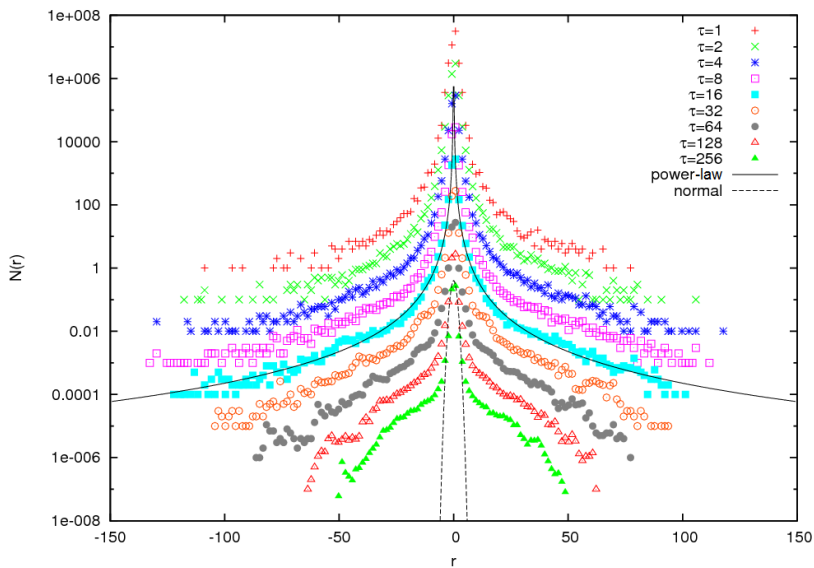}
\end{center}
\caption{Histograms of returns $r_{\tau}(t)$ from our simulation, for instance, for parameters $J=1, \lambda=1, K=5, b=2, b_0=0.21$, and different time lag $\tau$. The returns were re-scaled by the corresponding standard deviations of returns' time series. The distributions were separated for the better view. The dashed line is a normal distribution fitted to the central part of histogram for $\tau = 256$. The solid line denotes the power-law with exponent equals $3.322 = 1+\beta$ (see text for details).}
\label{stat}
\end{figure}

In Fig. \ref{stat} we plotted returns' histograms obtained in our simulation (in lin-log scale) for different values of time lag $\tau$.
In the simulation we used assumptions considered in Section \ref{SH}. Besides, we set, for instance, $J = 1, \lambda = 1, K = 5, b = 2, b_0 = 0.21$. Hence and from Formula (\ref{sigma2}), we obtain $\sigma = 0.42$ and $\alpha_{\sigma} = 0.105$ while $\alpha_{\lambda} = 0.25$. We verified that for assumed parameters only components with $j = 0,1,2,3,4$ are significant values in the sum in Ex\-pression (\ref{WM}). Apparently, the histograms for the most (i.e. for small and inter\-mediate) values of $\tau$ have well-formed power-law tails with exponent which (to good approximation) equals $1+\beta$, where $\beta=\ln{K}/\ln{b} = \ln{5}/\ln{2} = 2.322$. That is, these histograms quite well reproduce statistics given by Equation (\ref{appWM}). Remarkably, although the noise is cut off by borders $-(4J+\lambda)$ and $4J+\lambda$, the histograms of returns have still power-law tails driven by exponent $1 + \beta$. The more so, $\alpha_{\lambda} < 1$, which means that the range $[-(4J+\lambda), 4J+\lambda]$ is narrower than the corresponding one in the SH model. Nevertheless, for the largest values of $\tau$ the returns' distributions still exhibit fat tails. These results distinguish our approach from the SH model.

In addition we explain herein how we avoid a possible trapping in our simulation. By trapping we understand such a single realization of the simulation, where after sufficiently long time all spins are spontaneously ordered (a fully ferromagnetic state). Then the market is trapped by a stable state of extreme magnetization (except the case of $\alpha_{\lambda} > 1$ or $\alpha_{\sigma} > 1$ where this trapping does not occur at all). To avoid this trapping the system was activated by some exogenous factor making an abrupt transition of the system to the paramagnetic state\footnote{Abrupt transitions are a characteristic feature of modern financial markets.} (where all spins are randomly oriented). Then, the analysis of the system between these subsequent transitions is made.

\section{Conclusions}

In this work we essentially reinterpreted the Sieczka-Hołyst threshold model of financial markets. This reinterpretation means that we introduced a new meaning of the spin variable, although the mathematics of the SH model remained unchanged. In our approach the value of spin variable does not mean an action of the agent but, instead, it is only the agent's opinion about a market situation. This opinion he communicates to his neighbors. The agent's action (i.e. buying, selling or staying inactive) is in our approach connected only with the \emph{change} of spin variable. Indeed, this protects the stock market dynamics against the paradox (present in the SH model), where all agents buy stocks while corresponding temporal prices remain still unchanged.

Furthermore, our approach differs from the SH model as we used the Weierstrass-Mandelbrot noise to describe an intrinsic (erratic or even very intensive emotional) opinions of agents instead of the Gaussian noise used in the SH model.

Despite using of a threshold, the results obtained for Gaussian and \linebreak Weierstrass-Mandelbrot noises are quite different. The difference between both noises within the range $[-(4J + \lambda), 4J + \lambda]$ has a significant impact on the results. For instance, for the Weierstrass-Mandelbrot distribution there are power-law statistics for (almost) full range of returns, while for the Gaussian one the corresponding range is too short to say something definite. Obviously, the discreteness of Weierstrass-Mandelbrot distribution influences the results.

We simulated the above sketched dynamics of agents and calculated statistics (histograms) of returns. Some of these statistics have power-law tails while others have only fat tails. Explanation of this result is still a challenge. This is because power-law tails of the noise is cut-off by borders equal to $\pm(4J+\lambda)$. Nevertheless, power-law tails of returns' statistics is driven by the same exponent $\beta$ which governs the noise distribution. The systematic comparison of our approach predictions (additionally concerning e.g. autocorrelation functions and power spectrum) with empirical data would be a decisive verification of this approach\footnote{Our very initial comparisons give a promising agreement between predictions of our simulations and the corresponding empirical data \cite{DGK}.}.

\bibliographystyle{plainnat}

\end{document}